\documentclass[epjST]{svjour}
\usepackage{selinput}
\SelectInputMappings{
adieresis={ä},
germandbls={ß},
}
\usepackage{graphicx}
\usepackage{hyperref}
\usepackage{amssymb}
\usepackage{amsmath}
\usepackage{bbm}
\usepackage{color} 
\usepackage{rotating}

\newcommand{\bblinclude}{}

\renewcommand{\ProcessRunnHead}{
  \markboth{header to be inserted by the editor}{header to be inserted by the editor}}%

\newcommand{\trR}{\mathop{\mathrm{tr}_{\mathrm{R}}}}

\renewcommand{\phi}{\varphi}

\newcommand{\nc}[2]{\langle{#1}(t){#2}(t')\rangle}
\newcommand{\Si}{\mathsf{\Sigma}}
\renewcommand{\S}{\Sigma}

\title{A Variance Reduction Technique for the Stochastic Liouville-von Neumann
Equation}
\author{Konstantin Schmitz and Jürgen T. Stockburger\thanks{\email{juergen.stockburger@uni-ulm.de}}}

\institute{Institute for Complex Quantum
Systems, Ulm University, Ulm, Germany}

\date{2018-07-27}

\abstract{The Stochastic Liouville-von Neumann equation provides an
  exact numerical simulation strategy for quantum systems interacting
  with Gaussian reservoirs [J.T. Stockburger \& H. Grabert, PRL \textbf{88},
    170407 (2002)].  Its scaling with the extension of the time
  interval covered has recently improved dramatically through
  time-domain projection techniques [J.T. Stockburger, EPL \textbf{115}, 40010
    (2016)]. Here we present a sampling strategy which results in a
  significantly improved scaling with the strength of the dissipative
  interaction, based on reducing the non-unitary terms in sample propagation
  through convex optimization techniques.}
\begin{document}
\maketitle

\section{Introduction}

The notion of an open quantum system originates from the embedding of a system of interest into a larger environment, where system and environment together are governed by ordinary unitary quantum evolution. Different abstractions and techniques, both formal and computational have been put forward to formulate a reduced dynamics of the system of interest without explicit reference to the environment beyond its initial preparation.

Completely positive trace-preserving channels provide a very broadly applicable formal description of the time evolution of open quantum systems~\cite{alick87}. The proof of their existence is constructive; it does not, however, in itself provide a dynamical law governing the reduced description. For a quantum system with discrete energy spectrum, the respective quantum master equation can be derived within the constraints of a combined Born-Markov-secular approximation~\cite{davie74,breue02}.

The dissipator resulting from this depends not only on reservoir and coupling properties, but also on the energy spectrum and the eigenstates of the system. For complicated spectra, its determination can be cumbersome. The simpler alternative of adapting separate dissipators to individual parts of a system (``local'' Lindblad operators) may result in dynamics which contradicts elementary properties of thermal environments~\cite{levy14,stock17}, e.g., unphysical heat flow against a thermal gradient.
Similar problems arise in the context of driven quantum systems~\cite{alick06,schmi11}.

The formal description of environmental effect through influence functionals~\cite{feynm63,weiss08b} has the attractive feature of describing a dissipation mechanism  without direct reference to any system properties beyond specifying the coupling Hamiltonian. It is fully applicable in the context of any driving. In fact, it is completely ``agnostic'' to the nature of the free system dynamics, since it keeps the given separable structure of the interaction Lagrangian. Direct Monte Carlo evaluation of real-time path integrals is feasible~\cite{egger00,muhlb04} but numerically expensive in regimes without rapid dephasing by the environment (dynamical sign problem).  Equations of motion equivalent to the influence functional approach can be found only at the price of introducing a large hierarchy of auxiliary states~\cite{tanim91,tanim14}, since the weights assigned to paths by influence functionals are non-local in time.

On the other hand, the stochastic unraveling of the influence functional can restore this locality, at least on the level of individual realizations of auxiliary trajectories (which can be interpreted as Gaussian colored noise):
The Feynman-Vernon influence functional~\cite{feynm63} has essentially the same formal structure as the generating functional of Gaussian noise, hence it can be viewed as the noise average over exponentials of a time-local action functional~\cite{stock02}. This idea has led to a variety of stochastic propagation schemes~\cite{cao96,strun96,stock02,shao04,tanim06}. The corresponding numerical methods use independently drawn samples of the auxiliary Gaussian noise trajectories. For each noise sample, the time evolution of the reduced density matrix is obtained by integration of a Liouville equation with random terms, in a similar manner as a direct simulation of a generalized Langevin equation. The physical density matrix of the system is recovered by taking the average of these \emph{sample density matrices} in the limit of a large number of samples. It is to be noted that the sample density matrices are not physical quantum states; only their expectation value with respect to the probability measure of the Gaussian noise is to be identified with the reduced density matrix of the system.

The complete unraveling of Gaussian influence functionals introduced by Stockburger and Grabert~\cite{stock02} is valid for arbitrary spectral properties and temperatures of the reservoir. Its direct practical
application is sometimes hampered by rapid growth of the sample variance when long propagation intervals are covered. This problem has recently been solved for the near-universal case of a finite (not necessarily small) correlation time of the free reservoir fluctuations~\cite{stock16a}. The required number of samples has thus been reduced -- by orders of magnitude in some parameter regimes. For combinations of moderate to strong coupling and long reservoir memory times, the number of samples may be sufficiently high to require the use of parallel computational resources.

The present paper addresses this remaining problem by optimization of the noise correlation functions, which are not entirely determined by the unraveling procedure. Section \ref{sec:unrav} gives a brief overview of the unraveling of propagating functions based on influence functionals and their transformation into a stochastic Liouville-von Neumann (SLN) equation, including a discussion of issues related to sampling statistics. In section \ref{sec:opt} a strategy is developed which improves sampling statistics through minimizing the power of problematic noise components, and numerical examples are provided. Section \ref{sec:conc} comprises conclusions and an outlook.

\section{Stochastic unraveling and SLN equation}\label{sec:unrav}

The interacting dynamics of a system S coupled to an environment (reservoir) R is governed by a Hamiltonian which can be partitioned as $H = H_S + H_R + H_I$, where the indices S and R stand for terms which act exclusively on system and reservoir degrees of freedom. For the interaction we assume separability, $H_I = - A\otimes B$, where $A$ and $B$ act on system and reservoir.

We choose an initially thermal state of the environment (uncorrelated with the system) with the intent of tracing out the reservoir degrees of freedom from the correlated state arising from the dynamics. The unitary propagation of the global system-plus-reservoir density matrix is the appropriate conceptual starting point for this approach. Instead of using path integrals for this purpose~\cite{stock02}, equivalent time-ordered exponentials will be considered here. These can be considered generating functionals of non-commuting random variables, and they simplify to Gaussian expressions in exactly the same cases for which one obtains Gaussian influence functionals~\cite{kubo62}.

A most compact form of the dynamics \cite{hbarone} is obtained in the interaction representation of the Liouville dynamics,
\begin{equation}
\label{eq:dotrhoint}
\dot{\rho}(t) = i {A_+(t) B_-(t)} \rho(t) +i {A_-(t) B_+(t)} \rho(t) .
\end{equation}
For reasons which will become clear later, a choice was made to
decompose the interaction Liouvillian into superoperators for
reservoir $B_{\pm}$, and system $A_{\pm}$ defined through
(anti-)commutators, $B_- = [B,\cdot]$ and $B_+ = \frac{1}{2}
\{B,\cdot\}$ etc.

The formal solution of this dynamics is the time-ordered exponential
\begin{equation}
  \rho(t) =
  \exp_> \left(i \int_0^t ds
  ({A_+(s)\otimes B_-(s)} + {A_-(s)\otimes B_+(s)})
  \right)
  \rho(0).
\end{equation}

We now consider the case of Gaussian statistics of $B$ and an initially factorizing state, $\rho_S(0)\otimes \rho_R(0)$. Using the notation $\langle\cdot\rangle = \trR \left\{\cdot \rho_R(0)\right\}$ for the partial-trace averaging procedure, the \emph{reduced} density matrix is now formally given by
\begin{align}
  \rho_S(t) &=
  \left\langle\exp_> \left(i \int_0^t ds
  (A_+(s) B_-(s) \rho(s) + A_-(s) B_+(s))  
  \right)\right\rangle
  \rho_S(0) \nonumber\\
  &=
  \exp_> \bigg(i \int_0^t ds \int_0^s ds'
  (A_-(s)A_-(s')\big\langle B_+(s)B_+(s')\big\rangle\nonumber\\
  &\qquad\qquad\qquad\qquad\qquad+
  A_-(s)A_+(s')\big\langle B_+(s)B_-(s')\big\rangle )
  \bigg)\rho_S(0). \label{eq:gaussformal}
\end{align}
Of the four terms formally arising in a bivariate Gaussian characteristic function, two are identically zero here because they are traces over a commutator. A key observation at this point is the following: Had we started with a different averaging procedure -- replacing $B_+$ and $B_-$ by $c$-number Gaussian noise -- we would have arrived at a similar expression.

\begin{table}
\begin{tabbing}
state before reductionX \= $\rho$ in global Liouville spaceX
 \= \emph{random} $\rho$ in system space\kill\\
\> \underline{trace reduction}\> \underline{stochastic reduction}\\[1.05\baselineskip]
state before reduction:
 \> $\rho$ in global Liouville space \> \emph{random} $\rho$ in system space\\[0.8\baselineskip]
reservoir ``forces'': \> $B_+$, $B_-$ superoperators  \> $\xi(t)$, $\nu(t)$ random $\in \mathbbm{C}$\\[0.8\baselineskip]
reduction operation: \> partial trace $\mathrm{tr}_R$ \> expectation value\\
\> with weight $\rho_R$\> w.r.t. noise probability density\\[0.8\baselineskip]
\> $\langle B_+(t)B_+(t')\rangle$ \> $\langle \xi(t)\xi(t')\rangle$\\
\> $\langle B_+(t)B_-(t')\rangle$ \> $\langle \xi(t)\nu(t')\rangle$\\
\> $\langle B_-(t)B_+(t')\rangle=0$ \> $\langle \nu(t)\xi(t')\rangle=0$\\
\> $\langle B_-(t)B_-(t')\rangle=0$ \> $\langle\nu(t)\nu(t')\rangle=0$\\[0.5\baselineskip]
\> time ordering $t>t'$ \> time ordering ``by hand''
\end{tabbing}
\caption{Overview of the correspondence between Gaussian quantum fluctuations of an environment and stochastic processes with matched statistics.}
\label{tab:correspondence}
\end{table}

The conditions
\begin{align}
  \langle\xi(t)\xi(t')\rangle &= \langle B_+(t)B_+(t')\rangle
  = \Re \langle B(t)B(t')\rangle\label{eq:corrxixi}\\
  \langle\xi(t)\nu(t')\rangle &= \langle B_+(t)B_-(t')\rangle
  = {2i}\Theta(t-t') \Im\langle
B(t)B(t')\rangle \label{eq:corrxinu}\\
\langle\nu(t)\nu(t')\rangle &= 0, \label{eq:corrnunu}
\end{align}
are necessary and sufficient for the exact identification of the open-system dynamics (\ref{eq:gaussformal}) and stochastic propagation with the substitutions $B_+\to\xi$ and $B_-\to\nu$. On the left hand side, the angle brackets stand for expectation values with respect to the noise statistics. In practice, the stochastic equivalent of eq. (\ref{eq:gaussformal}) is evaluated by returning to the Schrödinger picture and averaging solutions of the SLN equation
\begin{equation}
  \dot{\tilde\rho} = \mathcal{L}_S \tilde\rho + i \xi(t) A_-\tilde\rho + i \nu(t) A_+\tilde\rho,
  \label{eq:SLN}
\end{equation}
where $\tilde\rho$ is the sample density matrix associated with any particular realization of $\xi(t)$ and $\nu(t)$.

Averaging $\tilde\rho$ over samples provides an estimate of the physical density matrix $\rho_S$. Since noise samples are generated independently, an accurate estimate of the statistical error of this approach can be determined. In this context, we also note that parallelization of the simulation method is trivial. An overview of the quantum-stochastic correspondence is given in Table 1.

It is now crucial to observe that
eqs. (\ref{eq:corrxixi}--\ref{eq:corrnunu}) cannot be obeyed by real-valued noise. Only when extending at least $\nu$ to values in the
complex plane can these equations hold: $\xi$ and $\nu$ are random
variables in the sense of ordinary Kolmogorov probability. When
extending both $\xi=\xi'+i\xi''$ and $\nu=\nu'+i\nu''$ to complex
values, the correlations of their four real components are not
completely determined by
eqs. (\ref{eq:corrxixi}--\ref{eq:corrnunu}). Additional non-physical
correlation functions, in complex notation,
$\langle\xi^*(t)\xi(t')\rangle$, $\langle\xi^*(t)\nu(t')\rangle$, and
$\langle\nu^*(t)\nu(t')\rangle$ may be modified without altering the
expectation value of samples propagated by
eq. (\ref{eq:SLN}). However, the generic constraint of non-negative
probability does apply and sets limits to the range of choices. Consequently, the solutions of eq. (\ref{eq:SLN}) are of the form $\tilde\rho(t) = R_L(t) \rho_S(0) R^\dagger_R(t)$, where $R_{L/R}(t)$ is almost surely non-unitary.

As a consequence of non-unitarity, the variance of samples, expressed by observables (or a suitable metric applied to $\tilde\rho$), tends to increase with the length of the propagation time interval. The typical asymptotic behavior, shared with many processes involving multiplicative noise~\cite{gardi09}, is exponential growth as a function of the interval length. Key quantities in this context are the noise spectra of $\nu''$ and $\xi''$, which drive both the sample trace and the Frobenius norm of $\tilde\rho$ towards a log-normal distribution. In some parameter regimes, this is harmless, in others, a finite-memory approach can halt this variance growth at a finite time~\cite{stock16a}. However, there is a problem in the strong-coupling limit, where the asymptotic dependence of sample variance on the coupling constant is exponential.

\section{Optimization of stochastic driving}\label{sec:opt}

Varying the non-physical correlations with the aim of minimizing $\nu''$ and $\xi''$ seems a natural desideratum. Formally, this is an optimization problem in a function space with explicit equality constraints and inequality constraints implied by the non-negativity of probability measures.

With the notable exception of Ref.~\cite{imai15}, this optimization was not attempted in previous work, and the non-physical correlations were chosen with an eye to easy numerical noise generation \cite{stock02,stock04,koch08}. $\xi$ was decomposed into a sum of independent terms $\xi^{l} \in \mathbbm{R}$ and $\xi^{s} \in \mathbbm{C}$ with
\begin{eqnarray}
  \label{eq:corrxilxil}
  \langle\xi^{(l)}(t)\xi^{(l)}(t')\rangle &=& \Re \langle B(t)B(t')\rangle \\
  \langle\xi^{(s)}(t)\nu(t')\rangle &=& {2i}\Theta(t-t') \Im\langle
B(t)B(t')\rangle + i\mu\delta(t-t'), \label{eq:crosscorrold}
\end{eqnarray}
where the prefactor $\mu$ of the Dirac delta function is chosen such that the time integral over (\ref{eq:crosscorrold}) vanishes~\cite{counterterm}. All other correlations are zero, except $\langle\nu^*(t)\nu(t')\rangle$ and $\langle(\xi^{(s)})^*(t)\xi^{(s)}(t')\rangle$, which are fixed by setting them equal to each other and assigning them the minimal autocorrelation noise power allowed by eq. (\ref{eq:crosscorrold}).

When minimizing the spectral noise power of $\nu''$ and $\xi''$, one needs to consider the full matrix of correlation functions after a Fourier transform,

\begin{equation}
  \Si(\omega) = F\left[\left(\begin{array}{cccc}
      \nc{\xi'}{\xi'} & \nc{\xi'}{\xi''} & \nc{\xi'}{\nu'} & \nc{\xi'}{\nu''}\\
      \nc{\xi''}{\xi'} & \nc{\xi''}{\xi''} & \nc{\xi''}{\nu'} & \nc{\xi''}{\nu''}\\
      \nc{\nu'}{\xi'} & \nc{\nu'}{\xi''} & \nc{\nu'}{\nu'} & \nc{\nu'}{\nu''}\\
      \nc{\nu''}{\xi'} & \nc{\nu''}{\xi''} & \nc{\nu''}{\nu'} & \nc{\nu''}{\nu''}
    \end{array}\right)\right]
\end{equation}
The rows and columns of $\Si$ will be naturally labeled by indices $\xi'$, $\xi''$, $\nu'$ and $\nu''$. With the notations $S(\omega) = F[\langle\xi(t)\xi(t')]$ and $D(\omega) = -i F[\langle\xi(t)\nu(t')]$, the operator-noise identifications (\ref{eq:corrxixi})--(\ref{eq:corrnunu}) now read
\begin{align}
  \S_{\xi'\xi'} + i \S_{\xi'\xi''} + i \S_{\xi''\xi'} - \S_{\xi''\xi''} &= S \label{eq:constraintxixi}\\
  \S_{\xi'\nu''} + \S_{\xi''\nu'} &= D\label{eq:constraintD}\\
  \S_{\xi'\nu'} + \S_{\xi''\nu''} &= 0\label{eq:constraintD0}\\
  \S_{\nu'\nu'} + i \S_{\nu'\nu''} + i \S_{\nu''\nu'} - \S_{\nu''\nu''} &= 0
  \label{eq:constraintnunu}
\end{align}
Because the spectra of real-valued correlation functions are considered at this point, pairs $(-\omega,\omega)$ must be considered in the optimization. As a consequence, condition (\ref{eq:corrxinu}) leads to two separate equations
(\ref{eq:constraintD}) and (\ref{eq:constraintD0}).

The condition of $\Si$ being positive semidefinite cannot easily be stated in the form of a simple algebraic expression. However, optimization problems involving semidefinite Hermitian matrices and linear constraints are special cases of convex optimization. These can be solved efficiently and reliably with modern numerical methods, implemented in packages such as CVX~\cite{grant14,grant08}, which we used to minimize $\S_{\xi''\xi''} + \S_{\nu''\nu''}$, subject to all of the constraints mentioned above.

The numerical solutions invariably show specific features which guide us towards an analytic solution:
\begin{enumerate}
\item $\Si$ is block diagonal; the pairs $(\xi',\nu'')$ and $(\xi'',\nu')$ are independent.
\item Each block is a rank 1 matrix.
\end{enumerate}
With these provisos, the solution of the optimization problem can be written in closed form,
\begin{align} \label{eq:optcorr}
  \left(\begin{array}{ll}
    \S_{\xi'\xi'} & \S_{\xi'\nu''}\\ \S_{\nu''\xi'} & \S_{\nu''\nu''}
  \end{array}\right)
  &=
  \left(\begin{array}{ll}
    A+S & D-C \\ D^*-C^* & B
  \end{array}\right)
  &,
  \left(\begin{array}{ll}
    \S_{\xi''\xi''} & \S_{\xi''\nu'}\\ \S_{\nu'\xi''} & \S_{\nu'\nu'}
  \end{array}\right)
  &=
  \left(\begin{array}{ll}
    A & C \\ C^* & B
  \end{array}\right)
\end{align}
with
\begin{align*}
A &= \frac{R^2}{1-2R} S
&
B &= (1-2R) |D|^2/S\\
C &= R D
&
R &= \frac{1}{2}\left(
1 - \frac{1}{\sqrt{4 |D|^2/S^2 + 1}}
\right).
\end{align*}
\begin{figure}
  \begin{center}
    \includegraphics[width=0.95\textwidth]{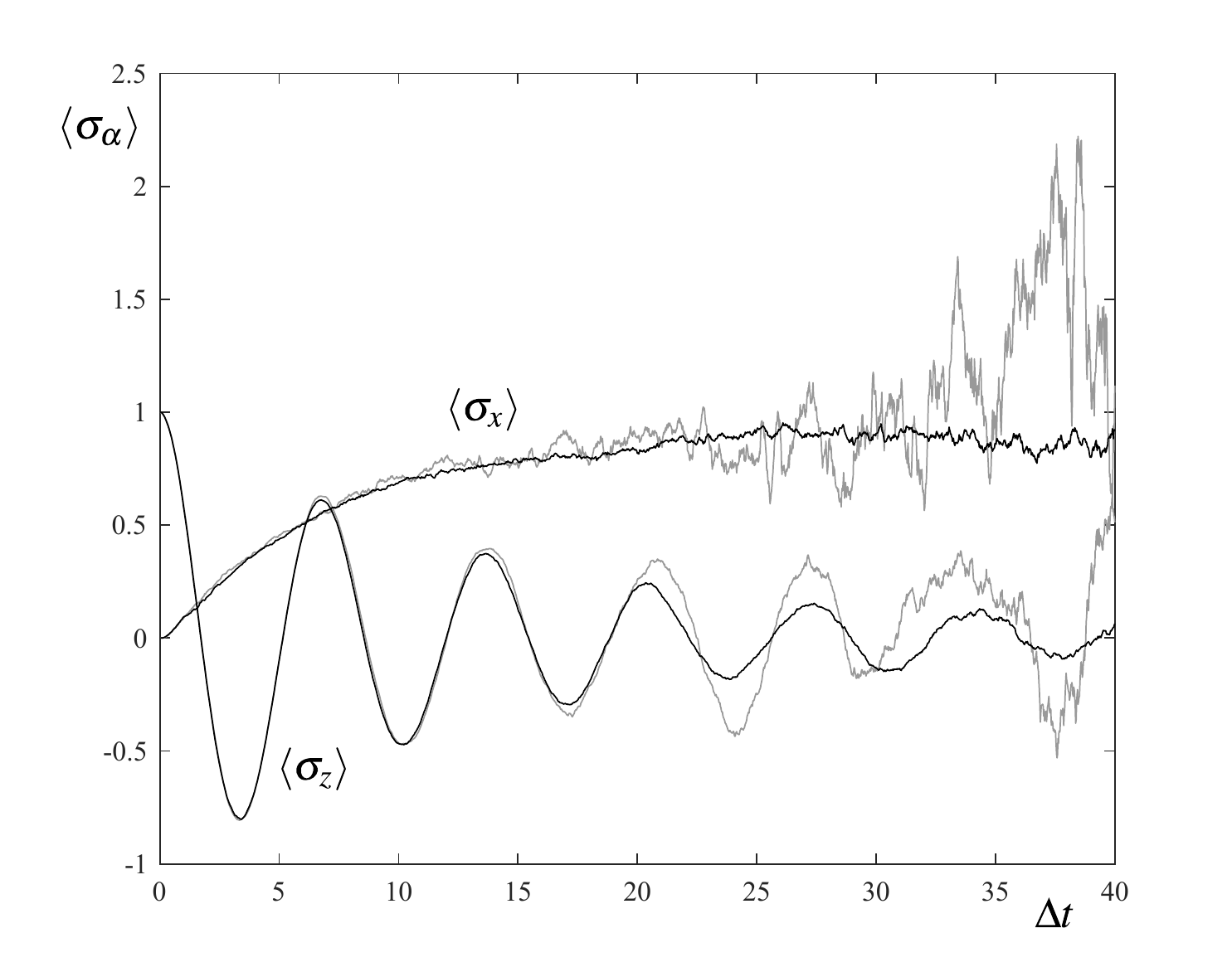}
  \end{center}
  \caption{Comparison of simulation data for the spin-boson system. Smoother lines (black) are with 5000 optimized noise samples; jagged lines (gray) with 5000 conventional samples. Parameters (with unit $\Delta=1$) are $\beta=5$, $\omega_c=10$ and dissipation constant $K=0.05$.}
  \label{fig:optresult}
\end{figure}

\begin{figure}
  \begin{center}
    \includegraphics[width=\textwidth]{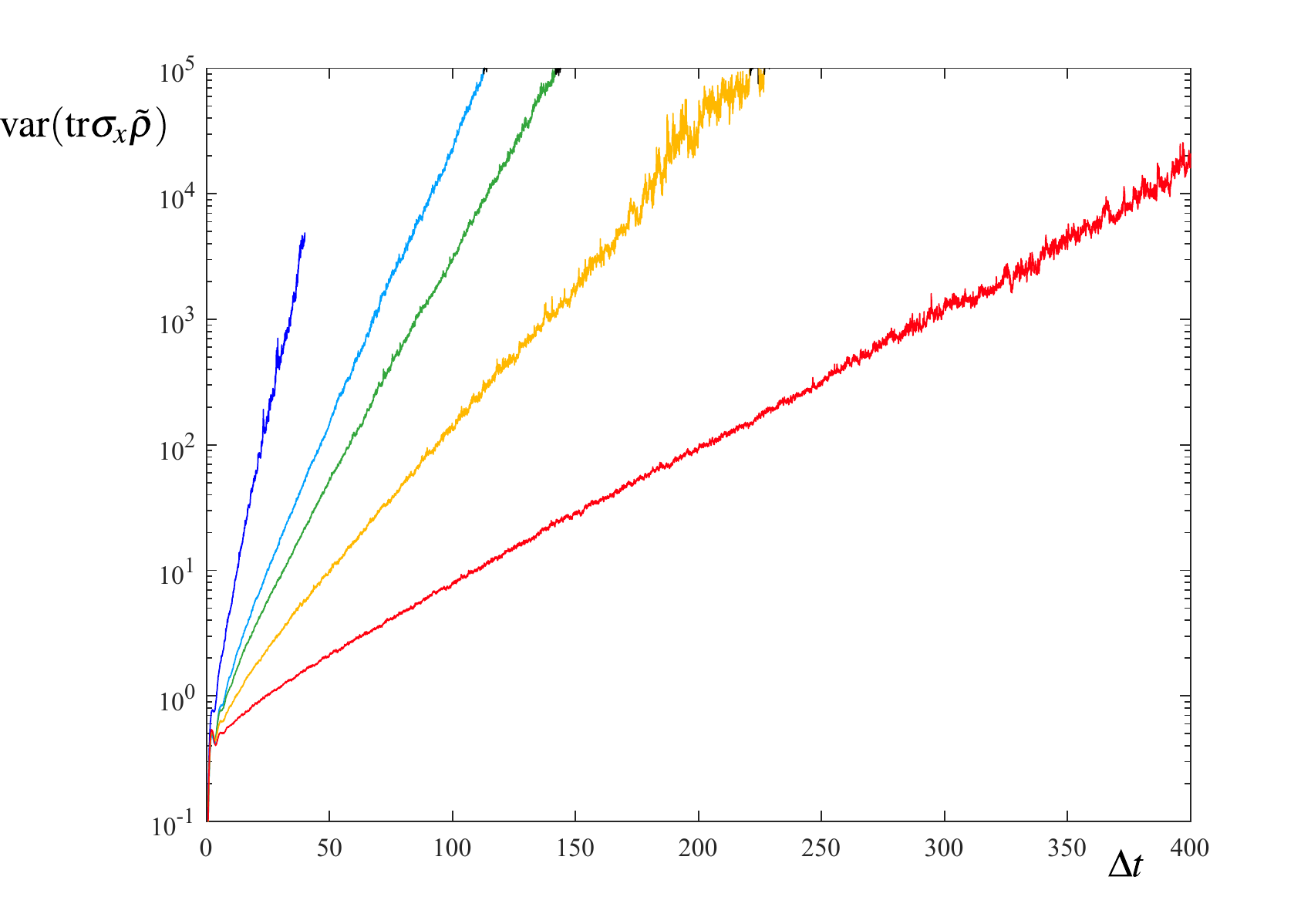}
  \end{center}
  \caption{(color online) Exponential increase of sample variance of the observable $\sigma_x$ with time. An excessively long propagation interval has been chosen to demonstrate exponential growth; the vertical line indicates $\Delta t=40$, a time at which all examples have come close to equilibrium. The uppermost curve is from the conventional simulation shown in Fig. \ref{fig:optresult} while the other four curves are from optimized simulations. Parameters are (unit $\Delta=1$) $\beta \in \{5,2,1,1/2\}$ (blue, light blue, green, yellow, red, top/left to bottom/right), $\omega_c=10$ and $K=0.05$.}
  \label{fig:vartime}
\end{figure}

For an Ohmic reservoir, the resulting noise powers $\S_{\xi''\xi''}$ and $\S_{\nu''\nu''}$ are always smaller than in the previously used construction using (\ref{eq:corrxilxil}) and (\ref{eq:crosscorrold}) by at least 30\%. The most significant advantage occurs in the case $|D|< S$, where the optimized values are $\S_{\nu''\nu''} \approx |D|^2/S$ and $\S_{\xi''\xi''} \approx |D|^4/S^3$, while the old approach had $\S_{\xi''\xi''} \approx \S_{\nu''\nu''} \approx |D|/2$. In the limit $|D|/S \to 0$, the optimal result coincides with the ansatz $\S_{\xi''\xi''}= 0$ made by Imai et al. for certain frequency ranges~\cite{imai15}.

In the context of Ohmic dissipation at finite temperature, the low-frequency behavior is quite relevant: It is a most important parameter determining the growth of the sampling variance discussed above. At frequencies below the thermal energy, the noise powers of $\xi''$ and $\nu''$ are now smaller by a factor of $\beta\omega$, i.e., the noise powers $\S_{\xi''\xi''}$ and $\S_{\nu''\nu''}$ now vanish quadratically in the infrared (instead of linearly). The comparison shown in Figures \ref{fig:optresult} and \ref{fig:vartime} demonstrates the extremely beneficial effect of this on the convergence of simulation data, using the spin-boson system~\cite{legge87} as an example. The symmetric spin-boson system is defined through $H_S = -\frac{\Delta}{2} \sigma_x$ and $A = \sigma_z$, using Pauli spin matrices. The reservoir is characterized by an Ohmic spectral density parameterized by UV cutoff frequency $\omega_c$ and a dimensionless dissipation constant $K$~\cite{weiss08b,stock04}. Figure \ref{fig:optresult} shows simulation results with identical parameters, including number of samples using the optimimzed noise statistics, eq. (\ref{eq:optcorr}), represented by smoother lines and the conventional statistics (\ref{eq:corrxilxil})--(\ref{eq:crosscorrold}), represented by lines indentifyable by visible growth of their noise amplitude.

Fig. \ref{fig:vartime} and table \ref{tab:growth} compare the performance characteristics of the old and new methods in further detail. Depending on temperature, the new approach reduces the growth rate of the sample variance significantly. The resulting savings factor in the required number of samples is therefore also an exponential function of simulation time. Savings factors of up to three orders of magnitude allow simulations extending up to approximate equilibration, using optimized samples. The savings factor is largest for high temperature and strong coupling. However, the asymptotic numerical cost still grows exponentially in the limit of very strong coupling.

\begin{table}[b]
  \begin{center}
  \newcommand\T{\rule{0pt}{2.6ex}}       
  \begin{tabular}{|l||c||c|c|c|c|} \hline
    \T & $\beta\Delta=5$ & $\beta\Delta=5$ & $\beta\Delta=2$ & $\beta\Delta=1$ & $\beta\Delta=0.5$ \\ \hline
    variance growth rate ($\Delta$) \T
    & $0.20$ & $0.10$
    & $8.6\cdot 10^{-2}$  & $5.5\cdot 10^{-2}$ & $2.5\cdot 10^{-2}$ \\ \hline
    variance at $\Delta t=40$ \T
    & $3\cdot 10^3$ & $5.2\cdot 10^2$
    & $1.8\cdot 10^1$ & $ 5.5 \cdot 10^0$ & $1.7\cdot 10^0$ \\ \hline
    extrapolated number  & & & & & \\
    of samples at $\Delta t=40$ &$3\cdot 10^7$ & $5.2\cdot 10^6$
    & $1.8\cdot 10^5$ & $5.5\cdot 10^4$ & $1.7\cdot 10^4$\\
    \hline
  \end{tabular}
  \end{center}
  \caption{Performance characteristics without (left of double line) and with noise optimization. The number of samples~\cite{nsampnote} results from demanding an absolute error of $0.01$ for the simulation result $\langle\sigma_x\rangle$. All data are for a symmetric spin-boson system with $K=0.05$.}
  \label{tab:growth}
\end{table}

\section{Conclusions}\label{sec:conc}
The noise samples used in an SLN-based propagation of open quantum systems can be optimized towards small imaginary parts. The resulting reduction in the number of required samples can be several orders of magnitude. No additional numerical costs arise from the optimization since it has an analytic solution. The range of applications for SLN simulations thus widens in virtually all parameter regimes, moderate or high temperature being the regime with the greatest benefit. With increasing availability of parallel computing resources, SLN-based methods are becoming an attractive, versatile tool for the dynamics of open quantum systems.\\[2\parskip]

\noindent The authors would like to thank two anonymous referees for valuable suggestions. K.S. showed that standard methods of semidefinite programming apply to the noise optimization and obtained the numerical optimization results; J.T.S. derived the analytic result (\ref{eq:optcorr}). Both authors contributed equally to the rest of the material.

\bibliographystyle{prsty}
\bibliography{FQMT17,endnotes}

\providecommand{\noopsort}[1]{}
\begin{thebibliography}{10}

\bibitem{alick87}
R. Alicki and K. Lendi, {\em Quantum Dynamical Semigroups and Applications},
  Vol.~286 of {\em Lecture Notes in Physics} (Springer, Berlin, 1987).

\bibitem{davie74}
E.~B. {Davies}, Communications in Mathematical Physics {\bf 39},  91  (1974).

\bibitem{breue02}
H.-P. Breuer and F. Petruccione, {\em The theory of open quantum systems}
  (Oxford University Press, Oxford, 2002), p.\ 625.

\bibitem{levy14}
A. {Levy} and R. {Kosloff}, EPL (Europhysics Letters) {\bf 107},  20004
  (2014).

\bibitem{stock17}
J.~T. Stockburger and T. Motz, Fortschritte der Physik {\bf 65},  1600067
  (2017).

\bibitem{alick06}
R. {Alicki}, D.~A. {Lidar}, and P. {Zanardi}, Phys. Rev. A {\bf 73},  052311
  (2006).

\bibitem{schmi11}
R. Schmidt {\it et~al.}, Phys. Rev. Lett. {\bf 107},  130404  (2011).

\bibitem{feynm63}
R.~P. Feynman and F.~L. Vernon, Ann. Phys. (N.Y.) {\bf 24},  118  (1963).

\bibitem{weiss08b}
U. Weiss, {\em Quantum dissipative systems}, No.~13 in {\em Series in modern
  condensed matter physics}, 3rd ed. (World Scientific, Singapore, 2008).

\bibitem{egger00}
R. Egger, L. M\"uhlbacher, and C.~H. Mak, Phys. Rev. E {\bf 61},  5961  (2000).

\bibitem{muhlb04}
L. M\"uhlbacher, J. Ankerhold, and C. Escher, J. Chem. Phys. {\bf 121},  12696
  (2004).

\bibitem{tanim91}
Y. Tanimura and P.~G. Wolynes, Phys. Rev. A {\bf 43},  4131  (1991).

\bibitem{tanim14}
Y. {Tanimura}, \jcp {\bf 141},  044114  (2014).

\bibitem{stock02}
J.~T. Stockburger and H. Grabert, Phys. Rev. Lett. {\bf 88},  170407  (2002).

\bibitem{cao96}
J. Cao, L.~W. Ungar, and G.~A. Voth, The Journal of Chemical Physics {\bf 104},
   4189  (1996).

\bibitem{strun96}
W.~T. Strunz, Phys. Lett. A {\bf 224},  25  (1996).

\bibitem{shao04}
J. Shao, J. Chem. Phys. {\bf 120},  5053  (2004).

\bibitem{tanim06}
Y. Tanimura, J. Phys. Soc. Jpn. {\bf 75},  082001  (2006).

\bibitem{stock16a}
J.~T. Stockburger, EPL (Europhysics Letters) {\bf 115},  40010  (2016).

\bibitem{kubo62}
R. Kubo, J. Phys. Soc. Jpn. {\bf 17},  1100  (1962).

\bibitem{hbarone}
For brevity, we use the convention $\hbar=1$ throughout the paper.

\bibitem{gardi09}
C.~W. Gardiner, {\em Stochastic methods: a handbook for the natural and social
  sciences}, No.~13 in {\em Springer series in synergetics}, 4th ed. ed.
  (Springer, Berlin, 2009).

\bibitem{imai15}
H. Imai, Y. Ohtsuki, and H. Kono, Chemical Physics {\bf 446},  134   (2015).

\bibitem{stock04}
J.~T. Stockburger, Chem. Phys. {\bf 296},  159  (2004).

\bibitem{koch08}
W. Koch, F. Gro\ss{}mann, J.~T. Stockburger, and J. Ankerhold, Phys. Rev. Lett.
  {\bf 100},  230402  (2008).

\bibitem{counterterm}
This must be compensated by a Hamiltonian term quadratic in $A$.

\bibitem{grant14}
M. Grant and S. Boyd, {CVX}: Matlab Software for Disciplined Convex
  Programming, version 2.1, \url{http://cvxr.com/cvx}, 2014.

\bibitem{grant08}
M. Grant and S. Boyd,  in {\em Recent Advances in Learning and Control}, {\em
  Lecture Notes in Control and Information Sciences}, edited by V. Blondel, S.
  Boyd, and H. Kimura (Springer, Berlin, 2008), pp.\ 95--110.

\bibitem{legge87}
A.~J. Leggett {\it et~al.}, Rev. Mod. Phys. {\bf 59},  1  (1987), {\em ibid.}
  {\bf 67}, 725 (1995) (erratum).

\bibitem{nsampnote}
Making use of a finite time window for reservoir correlations~\cite{stock16a},
  the absolute number of samples will be significantly smaller.

\end{thebibliography}
\bblinclude

\end{document}